\documentclass[preprint]{elsarticle}
\usepackage{lineno,hyperref}

\usepackage{graphicx}
\modulolinenumbers[5]
 \journal{J. Magn. Magn. Mater.}
\usepackage{dcolumn}
\usepackage{bm}
\usepackage{color}

\begin{document}

\begin{frontmatter}
\title{A first-principles DFT+$GW$ study of spin-filter and spin-gapless semiconducting Heusler compounds}

\author[UoP]{M. Tas}\ead{tasm236@gmail.com}

\author[Jul]{E. \c{S}a\c{s}{\i}o\u{g}lu\fnref{f1}}\ead{ersoy.sasioglu@physik.uni-halle.de}

\author[Jul]{C. Friedrich}

\author[UoP]{I. Galanakis}\ead{galanakis@upatras.gr}

\address[UoP]{Department of Materials Science, School
of Natural Sciences, University of Patras,  GR-26504 Patra,
Greece}

\address[Jul]{Peter Gr\"{u}nberg Institut and Institute for
Advanced Simulation, Forschungszentrum J\"{u}lich and JARA, 52425
J\"{u}lich, Germany}

\fntext[fn1]{Present address: Institut f\"ur Physik,
Martin-Luther-Universit\"at Halle-Wittenberg, D-06099 Halle
(Saale), Germany}

\begin{abstract}
Among Heusler compounds, the ones being magnetic semiconductors
(also known as spin-filter materials) are widely studied as they
offer novel functionalities in spintronic and magnetoelectronic
devices. The spin-gapless semiconductors are a special case. They
 possess a zero or almost-zero energy gap in one of the two
spin channels. We employ the $GW$ approximation to simulate the
electronic band structure of these materials. Our results suggest
that in most cases the use of $GW$ self energy instead of the
usual density functionals is important to accurately determine the
electronic properties of magnetic semiconductors.
\end{abstract}

\end{frontmatter}

\section{Introduction}

Spintronics and magnetoelectronics constitute one of the most
rapidly expanding research fields of materials science and
condensed matter physics \cite{ReviewSpin}. The on-going research
on the modelling of novel materials plays a key role in the
advancements in this research field as it allows an \`a-la-carte
design of materials for specific applications \cite{Sato}. In this
respect, Heusler compounds \cite{heusler,landolt,landolt2} are
widely studied due to the variety of magnetic properties
exhibited, in particular, in combination with the half-metallicity
\cite{Gillessen2008,Gillessen2009,Butler,Faleev,Katsnelson,Galanakis02,Wang1}.
Apart from half-metallic Heusler compounds, also the ones being
magnetic semiconductors are of interest for spintronics and
magnetoelectronics \cite{book}. Magnetic semiconductors can act as
spin-filter materials (SFMs) \cite{ReviewSPT,ReviewSPT2} to
maximize the efficiency of devices based on magnetic tunnel
junctions (MTJs) \cite{LeClair02,Filip02,HybridMTJ,Muller}, like
the recently proposed spin-current diodes \cite{Sun}. A special
class of magnetic semiconductors are the so-called spin-gapless
semiconductors (SGSs), where one of the two spin channels presents
a gapless or almost-gapless semiconducting behavior while the
other spin channel possesses a finite gap at the Fermi level
\cite{Lukashev,Ouardi,Wang2}.

Several studies have been devoted recently to the SFMs and SGSs.
First-principles electronic band structure calculations have
suggested that the ordered quaternary (CoV)YAl and (CrV)YAl
Heusler compounds \cite{Galanakis13b,Alijani}  - where Y stands
for Ti, Zr, or Hf - are SFMs \cite{SFM,SFM2,SFM3,GalaRev}. The
former compounds are ferromagnetic semiconductors, while the
latter ones are fully-compensated ferrimagnetic semiconductors, as
they combine the existence of energy gaps in both spin directions
to zero magnetization \cite{SFM,SFM2}. Both families of Heusler
SFMs present high Curie temperatures well above the room
temperature and, thus, are of interest for room-temperature
spintronic/magnetoelectronic applications contrary to other well
known SFMs \cite{Review_EuChalcogenides,EuO_EuSe,EuS,Nagahama07}.
The fully-compensated ferrimagnetic character of (CrV)YAl
compounds makes them even more attractive for applications as the
zero net magnetization leads to vanishing stray fields and thus to
minimal energy losses. Recently, Stephen and collaborators have
successfully grown samples of (CrV)TiAl and their measurements
confirmed the \textit{ab-initio} predictions \cite{CrVTiAl-Exp}.

SGS materials are known for almost a decade
\cite{Wang,Wang09,Kim2014,Choo}. Although several
\textit{ab-initio} calculations had suggested that the band
structure of Mn$_2$CoAl, an inverse Heusler compound \cite{GSP},
is compatible with that of an SGS material
\cite{Liu08,Meinert11,Meinert11c}, it was not until 2013 when
experiments by Ouardi \textit{et al.} confirmed the SGS character
of Mn$_2$CoAl in bulk-like polycrystalline films and measured a
Curie temperature of 720 K far above the room temperature
\cite{Ouardi}. Also, experiments were carried out on thin films of
Mn$_2$CoAl on various substrates; films on top of GaAs were found
to deviate from SGS \cite{Jamer13,Jamer14}, while films on
thermally oxidized Si substrates were found to be SGS with a Curie
temperature of 550 K \cite{Xu2014}. These experimental findings
were accompanied by several \textit{ab-initio} calculations
\cite{GalaRev,Galanakis13b,GalanakisSGS,Xu13,Gao13,Jia2014,Wollmann,Bainsla,Bainsla2},
and also Ti$_2$CoSi, Ti$_2$MnAl, Ti$_2$VAs and Cr$_2$ZnSi have
been identified as potential SGS materials sublattices
\cite{GalanakisSGS}. Simultaneously, several theoretical studies
appeared dealing with the various phenomena that affect the SGS
character of these compounds with a focus mainly on Mn$_2$CoAl
\cite{Galanakis2014,Zhang13,Kudrnovsky,Jakobsson}.

\begin{figure}
\begin{center}
\includegraphics[width=\columnwidth]{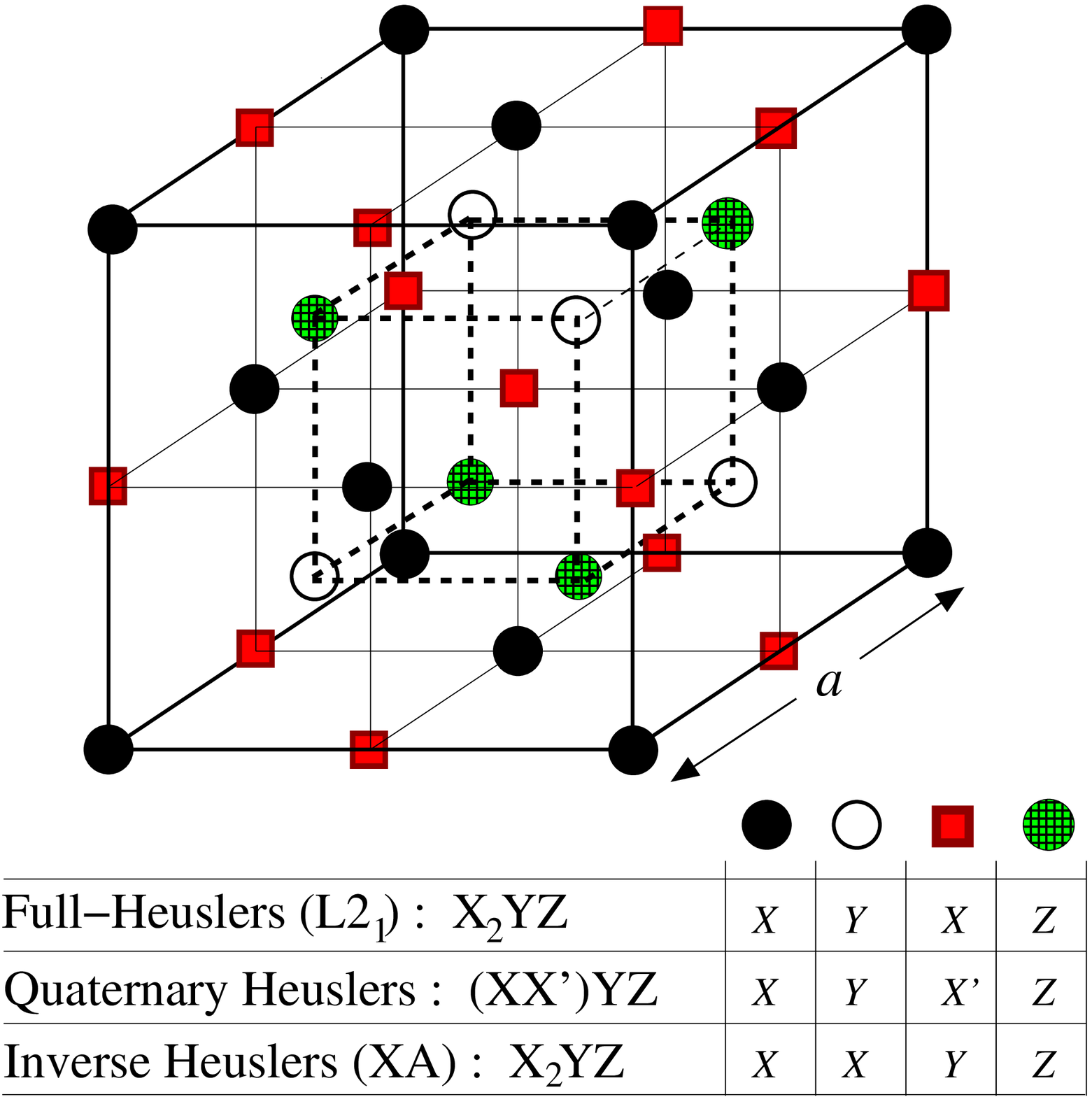}
\end{center}
\caption{(Color online) Schematic representation of the lattice
structure of the full-Heusler compounds adopting the $L2_1$
lattice structure, the quaternary ordered ones, which include also
the spin-filter materials, and the inverse Heusler compounds
adopting the $XA$ (also known as  $X_a$) lattice structure which
is identical to that of the spin-gapless semiconducting Heusler
compounds. In all cases, the lattice consists of four
interpenetrating fcc lattices. \label{fig1}}
\end{figure}

\section{Motivation and computational details}

For SGS materials the most important property is the zero gap in
the majority-spin or spin-up (for materials with zero
magnetization where we cannot distinguish majority and minority
spins) electronic band structure. In the case of SFMs the most
important property is the energy difference in the bottom of the
conduction band for the two spin directions, the exchange
splitting $2\Delta E_{\mathrm{ex}}$, which corresponds to the
difference of the barrier that spin up and spin down electrons
confront when they tunnel through the SFM \cite{Santos}. This
exchange splitting for the SFMs under study is about 0.1-0.3 eV
\cite{SFM,SFM2}. Thus, the energy quantities characterizing the
SGSs and SFMs are relatively small.

\begin{table*}
\caption{For the spin-filter and spin-gapless semiconducting
materials we present the used theoretical lattice constants $a$
(in \AA) from references \cite{SFM,SFM2,GalanakisSGS} with the
exception of Mn$_2$CoAl where we have used the experimental value
from reference \cite{Ouardi}, the magnetic energy $\Delta E_M$ in
eV defined as the difference between the magnetic and the
non-magnetic calculations, the atom-resolved spin magnetic moments
(in $\mu_\mathrm{B}$), the total spin magnetic moment
$m^\textrm{f.u.}$ (in $\mu_\mathrm{B}$),  and the exchange
splitting $2\Delta E_{\mathrm{ex}}$ (in eV) obtained from both the
PBE functional and the $GW$ self-energy (all other quantities are
computed using PBE only). \label{table1}}
\begin{tabular}{lcc|ccccc|cc}\hline

(XX$^\prime$)YZ &   $a$(\AA) &  $\Delta E_M$ &  $m^\mathrm{X}$ &
  $m^\mathrm{X^\prime}$ &  $m^\mathrm{Y}$ &
$m^\mathrm{Z}$ &  $m^\mathrm{f.u.}$  &   2$\Delta
E_\mathrm{ex}^\mathrm{PBE}$  &   2$\Delta E_\mathrm{ex}^{GW}$
\\ \hline
(CoV)TiAl &  6.04  &   -0.66 &   0.249  &   2.162 &  0.279 &  0.023 & 3.0    & 0.04 & 0.14     \\
(CoV)ZrAl &  6.26  &   -0.77 &   0.147  &   2.329 &  0.168 &  0.018 & 3.0    & 0.25 & 0.18     \\
(CrV)TiAl &  6.20  &   -0.78 &   -2.740 &   2.092 &  0.417 &  0.023 & 0.0    & 0.29 & 0.28       \\
(CrV)ZrAl &  6.41  &   -0.91 &   -2.913 &   2.362 &  0.251 &  0.043 & 0.0    & 0.27 & 0.26       \\
\hline X$_2$YZ &   $a$(\AA) &  $\Delta E_M$ & $m^{\mathrm{X}^A}$ &
  $m^{\mathrm{X}^B}$ &  $m^\mathrm{Y}$ &
$m^\mathrm{Z}$ &  $m^\mathrm{f.u.}$  &   2$\Delta
E_\mathrm{ex}^\mathrm{PBE}$  &   2$\Delta E_\mathrm{ex}^{GW}$
\\ \hline
Cr$_2$ZnSi &   5.85  &  -0.11  &   -1.58 &   1.62 &   0.03    &   -0.06 & 0.0   & 0.28 & 0.52  \\
Mn$_2$CoAl &   5.798 &  -0.95  &   -1.72 &   2.75 &   1.04    &   -0.04 & 2.0   & 0.34 & 0.36  \\
Ti$_2$CoSi &   6.03  &  -0.36  &   1.41  &   0.71 &   0.40    &   0.05  & 3.0   & 0.11 & 0.02  \\
Ti$_2$MnAl &   6.24  &  -0.30  &   1.13  &   1.00 &   -2.59 &
0.04  & 0.0   & 0.43 & 0.42  \\   \hline
\end{tabular}
\end{table*}

The fact that we have to deal with such small energy differences
calls for a sophisticated theoretical method for the description
of the electronic excitation spectrum or, in other words, the band
structure of these materials. While being very successful for
ground-state properties, standard DFT functionals are not adequate
for this purpose. This is not a failure of the approximations but
rather because the Kohn-Sham eigenvalues are not meant to be
interpreted as the excitation energies of the real interacting
system. Technically, these eigenvalues are the excitation energies
of an unphysical, auxiliary system of non-interacting electrons,
the Kohn-Sham system. As a consequence, they miss renormalization
effects due to electronic exchange and correlation effects.
Therefore, we employ in the present study the $GW$ approximation
 for the electronic self-energy, which is derived in
the framework of many-body perturbation theory and, thus, treats
the interactions among the electrons beyond the mean-field
approximation \cite{Hedin}. It contains the electronic exchange
exactly and a large part of electronic correlation. This approach
is well known to have a strong effect on the band gaps of
semiconductors and insulators \cite{Aulbur}. In particular, the
$GW$ approach corrects the band gaps from their (usually
underestimated) DFT values towards experiment. Furthermore, it is
known to produce more accurate results for half-metallic Heusler
compounds than other simplified approaches such as GGA+$U$
\cite{GalaU1,GalaU2}. For the half-metallic Heusler compounds
Co$_2$MnSi and Co$_2$FeSi, it has also been shown to be able to
accurately reproduce the experimental photoemission and x-ray
absorption spectra \cite{Meinert}. Thus, we expect the $GW$
self-energy to play an important role in the theoretical
description of the SFM and SGs materials; this assumption is
confirmed by our results in the next two sections.

We have chosen as SFMs the ferromagnetic (CoV)TiAl and (CoV)ZrAl
semiconductors \cite{SFM} and the fully-compensated ferrimagnetic
(CrV)TiAl and (CrV)ZrAl semiconductors \cite{SFM2}. As SGS
materials to study, we have chosen Mn$_2$CoAl and Ti$_2$CoSi which
present a ferrimagnetic and ferromagnetic configuration,
respectively, with a non-zero net magnetization, as well as
Cr$_2$ZnSi and Ti$_2$MnAl which are fully-compensated ferrimagnets
\cite{GalanakisSGS}. The lattice structure of all compounds under
study is presented in figure \ref{fig1}. The lattice has a fcc
structure with four atoms as basis along the diagonal. In the
usual full-Heusler compounds having the chemical formula X$_2$YZ,
the sequence of the atoms is X-Y-X-Z. For the quaternary Heuslers
like the SFMs under study, having the chemical formula
(XX$^\prime$)YZ the sequence of the atoms is X-Y-X$^\prime$-Z.
Finally, for the so-called inverse Heuslers, which have the same
chemical formula as the usual full-Heuslers with a larger valence
on Y than on X, like in the studied SGS, the sequence of the atoms
changes and it is now X-X-Y-Z (we use the superscripts A and B to
distinguish the two non-equivalent X atoms). For all cases we have
used the theoretical equilibrium lattice constants determined in
references \cite{SFM,SFM2,GalanakisSGS} and shown in table
\ref{table1}, with the exception of Mn$_2$CoAl where we have used
the experimental lattice constant in reference \cite{Ouardi}.

As a first step we performed simulations using the standard
density-functional-theory (DFT) based on the full-potential
linearized augmented-plane-wave (FLAPW) method as implemented in
the \texttt{FLEUR} code \cite{Fleur} within the
generalized-gradient approximation (GGA) of the
exchange-correlation potential as parameterized by Perdew, Burke
and Ernzerhof (PBE) \cite{PBE}. By using the PBE results as an
input, we performed  calculations employing the $GW$ approximation
using the SPEX code \cite{SPEX}. Details of the calculations are
identical to the ones in references \cite{Tas,Tas2}, where
non-magnetic semiconducting and antiferromagnetic semiconducting
Heusler compounds were studied, respectively.

\begin{figure}
\begin{center}
\includegraphics[width=\columnwidth]{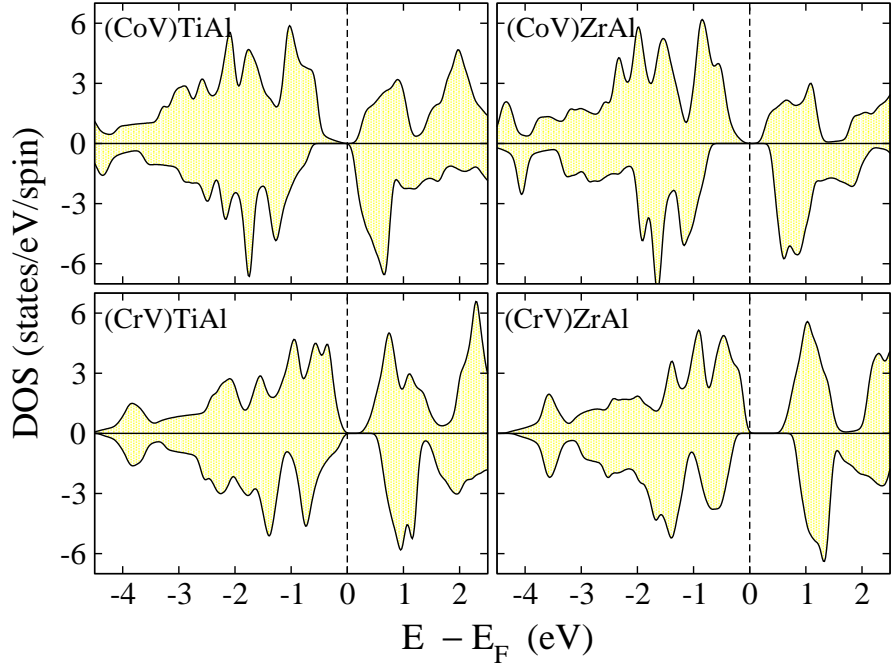}
\end{center} \caption{(Color online) Density of states for all four
SFMs under study obtained from the PBE  approximation. The zero
energy value denotes the Fermi level. Positive(negative) DOS
values correspond to the spin-up(spin-down)electrons.
 \label{fig2}}
\end{figure}

\section{Results on the spin-filter materials}

We begin our discussion with the SFMs. We have examined first, the
stability of the magnetic states by calculating the energy
difference between the magnetic and the non-magnetic states,
$\Delta E_M$, using the PBE functional, see table \ref{table1}.
For all four SFMs under study the obtained values are negative and
thus the magnetic state is favorable. Moreover, the calculated
values range between -0.66 eV for (CoV)TiAl down to -0.91 eV for
(CrV)ZrAl. These values are in perfect agreement with the
calculated values in references \cite{SFM} and \cite{SFM2} where a
different electronic structure method was employed, the
full-potential nonorthogonal local-orbital minimum-basis band
structure scheme (FPLO) \cite{koepernik}. Thus, these compounds
should exist at least as metastable structures as also suggested
by the experiment in \cite{CrVTiAl-Exp}.

Second, using the PBE functional, we have calculated  the
atom-resolved spin magnetic moments as well as the total spin
magnetic moment per formula unit (f.u.) which coincides with the
per unit cell value and we present these results also in table
\ref{table1}. For the two compounds containing Co, the total spin
magnetic moment is 3 $\mu_\mathrm{B}$, while for the two Cr-based
compounds the calculated total spin magnetic moment is exactly
zero in agreement with the calculations in references \cite{SFM}
and \cite{SFM2}. The Ti atoms in all four compounds carry a small
spin magnetic moment being ferromagnetically coupled to the spin
magnetic moment of the V atoms which range from about 2.1
$\mu_\mathrm{B}$ to about 2.4 $\mu_\mathrm{B}$. In the case of
(CoV)TiAl and (CoV)ZrAl the spin magnetic moments of the Co atoms
are also ferromagnetically coupled to the spin magnetic moments of
the V atoms leading to a ferromagnetic state. On the contrary, in
the case of (CrV)TiAl and (CrV)ZrAl compounds the Cr atoms carry
very large negative spin magnetic moments approaching -3
$\mu_\mathrm{B}$, which balance the positive spin magnetic moments
of the other atoms leading to the fully-compensated ferrimagnetic
ground state. The origin of this coupling can be easily explained
using the phenomenological Bethe-Slater rule as discussed in
reference \cite{SFM2}. The atom-resolved spin magnetic moments
here are also similar to the values calculated in references
\cite{SFM} and \cite{SFM2}; the only noticeable difference being
that the absolute values of the spin magnetic moments are slightly
smaller.

\begin{figure}
\begin{center}
\includegraphics[width=\columnwidth]{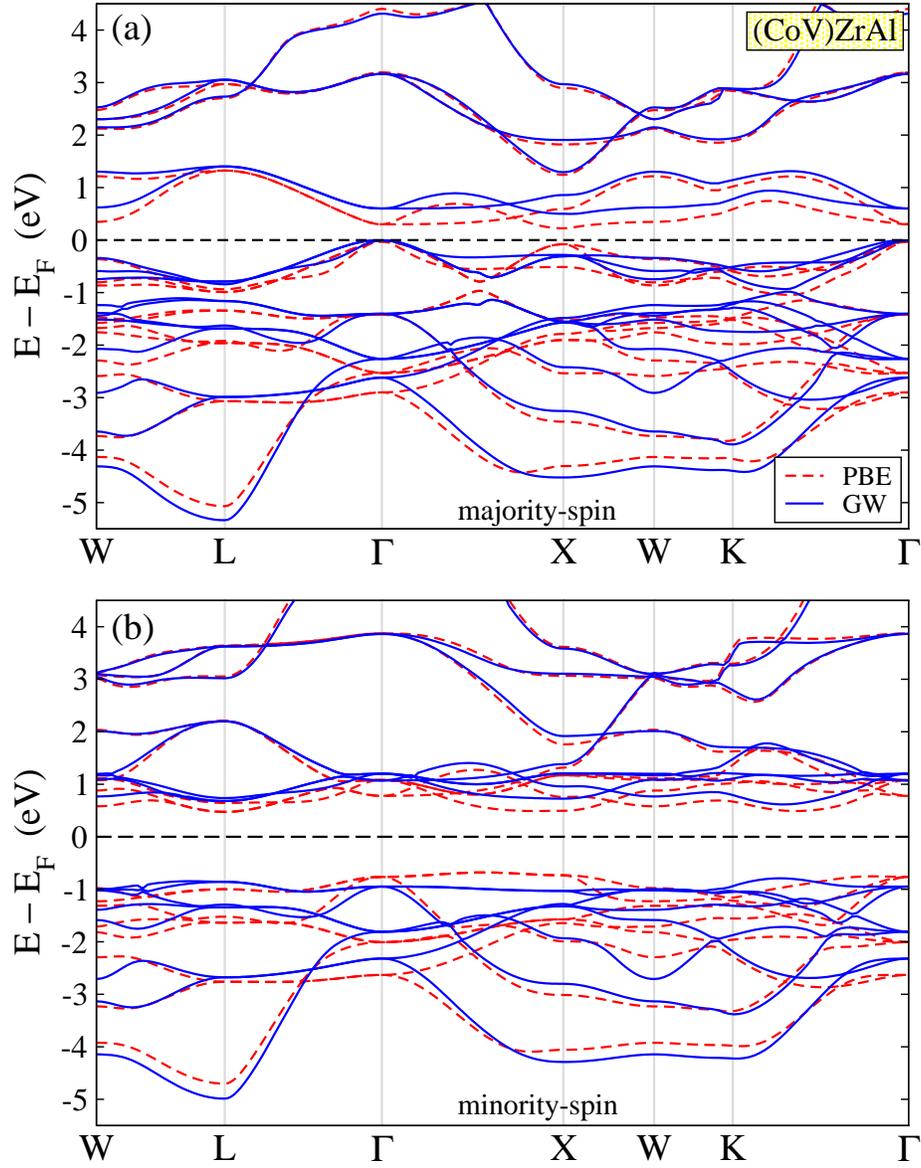}
\end{center} \caption{(Color online) Spin-resolved  electronic band
structure of  (CoV)TiAl along the high-symmetry directions in the
first Brillouin obtained from  PBE (red dashed line) and $GW$
(blue solid line) approximations. The zero energy value denotes
the Fermi level.
 \label{fig3}}
\end{figure}

\begin{figure}
\begin{center}
\includegraphics[width=\columnwidth]{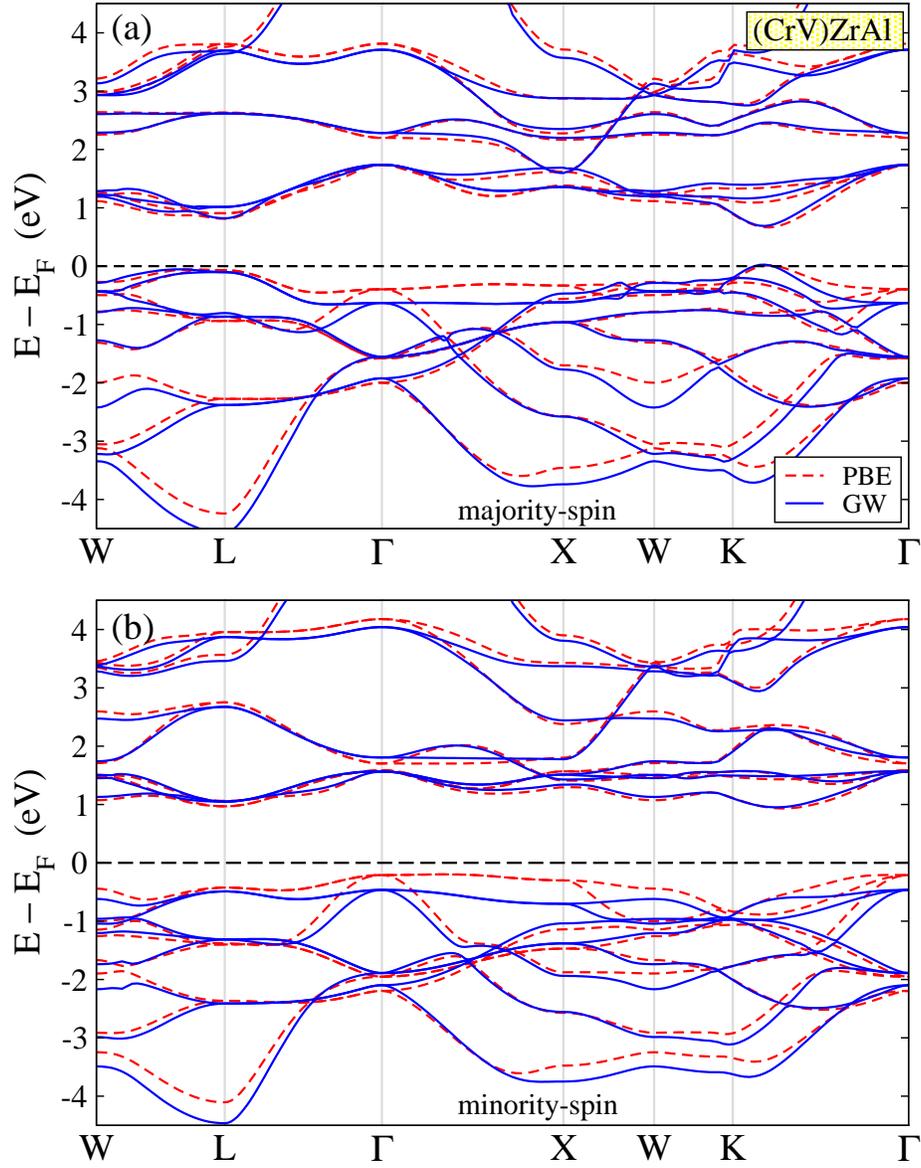}
\end{center} \caption{(Color online) Same as figure \ref{fig3} for (CrV)TiAl.
 \label{fig4}}
\end{figure}

The density of states (DOS) shown in figure \ref{fig2} for all
four SFMs under study is compatible with a magnetic semiconducting
ground state in agreement with the FPLO results in references
\cite{SFM} and \cite{SFM2}. In the last columns of table
\ref{table1} we have compiled the exchange splittings obtained
from PBE and $GW$. For (CrV)TiAl and (CrV)ZrAl both approximations
yield similar values with a difference of only 0.01 eV. Also the
calculated values for these two Cr-based compounds are identical
to the values of 0.28 and 0.25 eV calculated in reference
\cite{SFM2}. The discrepancy between the $GW$ and the PBE
calculations is larger for the case of (CoV)TiAl and (CoV)ZrAl
compounds. For the first compound $GW$ increases the exchange
splitting by 0.1 eV while for the second compound it decreases by
0.07 eV. We have to note here that the present calculations using
PBE yield, with respect to the PBE calculations in reference
\cite{SFM}, half the exchange splitting energy for (CoV)TiAl. On
the contrary, the present first-principles calculations using the
PBE functional yield for (CoV)ZrAl an exchange splitting identical
to the calculated value in reference \cite{SFM}. Our $GW$ results
suggest that (CrV)TiAl and (CrV)ZrAl are more suitable for
applications since the large values of the exchange splitting
ensure a more efficient spin-dependent tunnelling in realistic
devices \cite{ReviewSPT,ReviewSPT2}.

\begin{table*}
\caption{Calculated PBE (in parenthesis) and \textit{GW} energy
band gaps and transition energies (all in eV) between certain
high-symmetry points for  both spin channels of the spin-filter
materials under study. For the case of fully-compensated
ferrimagnets, where spin-up and spin-down bands have the same
population, we choose as majority-spin band structure the one with
the smallest energy band gap (E$_{\textrm{g}}$).\label{table2}}
\begin{tabular}{lccccccc}\hline
(XX$^\prime$)YZ &
E$_{\textrm{g}}^{GW}$(E$_{\textrm{g}}^\mathrm{PBE}$) & $\Gamma
\rightarrow \Gamma$ & X$\rightarrow$X & L$\rightarrow$L &
 $\Gamma \rightarrow$X & $\Gamma\rightarrow$L & X$\rightarrow$L \\
\hline
& \multicolumn{7}{c}{\underline{Majority-Spin Band-Structure}} \\
(CoV)TiAl &     0.44(0.23)&   0.51(0.31)&   1.21(0.74) &   2.28(2.24) &   0.44(0.23)&   1.29(1.21)&   2.06(1.71) \\
(CoV)ZrAl &     0.50(0.25)&   0.60(0.33)&   0.80(0.30) &   2.24(2.26) &   0.50(0.25)&   1.40(1.35)&   1.71(1.40) \\
(CrV)TiAl &     0.45(0.37)&   2.02(2.05)&   1.79(1.54) &   1.08(1.00) &   1.33(1.30)&   0.98(0.94)&   1.44(0.94) \\
(CrV)ZrAl &     0.66(0.66)&   2.37(2.13)&   1.97(1.71) &   0.92(0.88) &   1.98(1.78)&   1.45(1.21)&   1.44(1.44) \\
& \multicolumn{7}{c}{\underline{Minority-Spin Band-Structure}} \\
(CoV)TiAl &     0.96(0.80)&   1.63(1.34)&   2.06(1.51) &   1.60(1.48) &   1.43(1.21)&   1.09(0.88)&   1.73(1.18) \\
(CoV)ZrAl &     1.53(1.15)&   2.01(1.54)&   1.75(1.23) &   1.53(1.47) &   1.67(1.26)&   1.63(1.24)&   1.70(1.21) \\
(CrV)TiAl &     0.81(0.70)&   1.56(1.55)&   2.08(1.70) &   1.56(1.40) &   1.25(1.20)&   0.94(0.78)&   1.77(1.27) \\
(CrV)ZrAl &     1.42(1.14)&   2.02(1.80)&   2.04(1.60) &
1.54(1.39) &   1.80(1.50)&   1.51(1.18)&   1.75(1.27)
\\ \hline
\end{tabular}
\end{table*}

In figures \ref{fig3} and \ref{fig4} we present the spin-resolved
band structure for (CoV)TiAl and (CrV)TiAl using both the PBE
functional (red dashed line) and the $GW$approximation (solid blue
line). Note that for (CrV)TiAl both spin directions have the same
population of electrons and thus arbitrarily we denote as
majority-spin band the one with the smallest energy band gap which
corresponds to the negative spin magnetic moment of the Cr atoms
in table \ref{table1}. The Fermi level is set to be the top of the
valence band of the majority-spin electrons. While for the Heusler
compounds studied in references \cite{Tas} and \cite{Tas2} $GW$
only marginally affected the bands close to the Fermi level, in
the case of SFMs $GW$ has a more profound effect on the bands just
above and below the Fermi level. Most of the bands around the
Fermi level are shifted away from the Fermi level leading to
larger band gaps. This shift is quantified for all compounds under
study in the first column of table \ref{table2}, where we present
for all four compounds and for both spin directions the band gap
using both the $GW$ approximation and the PBE functional in
parenthesis. The largest change in percentage is observed for the
majority-spin band of (CoV)ZrAl where the band gap within $GW$ is
doubled from 0.25 eV to 0.50 eV, and the largest change in
absolute values is in the minority spin band structure of the same
compound where $GW$ increases the gap by 0.38 eV from 1.15 eV to
1.53 eV. The only exception in the behavior is the majority spin
band structure of (CrV)ZrAl where the band gap is 0.66 eV using
both $GW$ and PBE. The effect of the $GW$ renormalization is even
more important for the transition energies shown in table
\ref{table2} as well. For example, in the majority spin band
structure of (CoV)ZrAl the \textbf{k}-conserved transition energy
from the valence to the conduction band at the X point is 0.30 eV
in PBE, but it increases to 0.80 eV using the $GW$ approximation.
Thus, one may conclude that correlations play an important role in
spin-filter Heusler compounds and the use of more sophisticated
schemes  than the simple density functionals is needed to
correctly describe their electronic band structure properties.

\begin{figure}
\begin{center}
\includegraphics[width=\columnwidth]{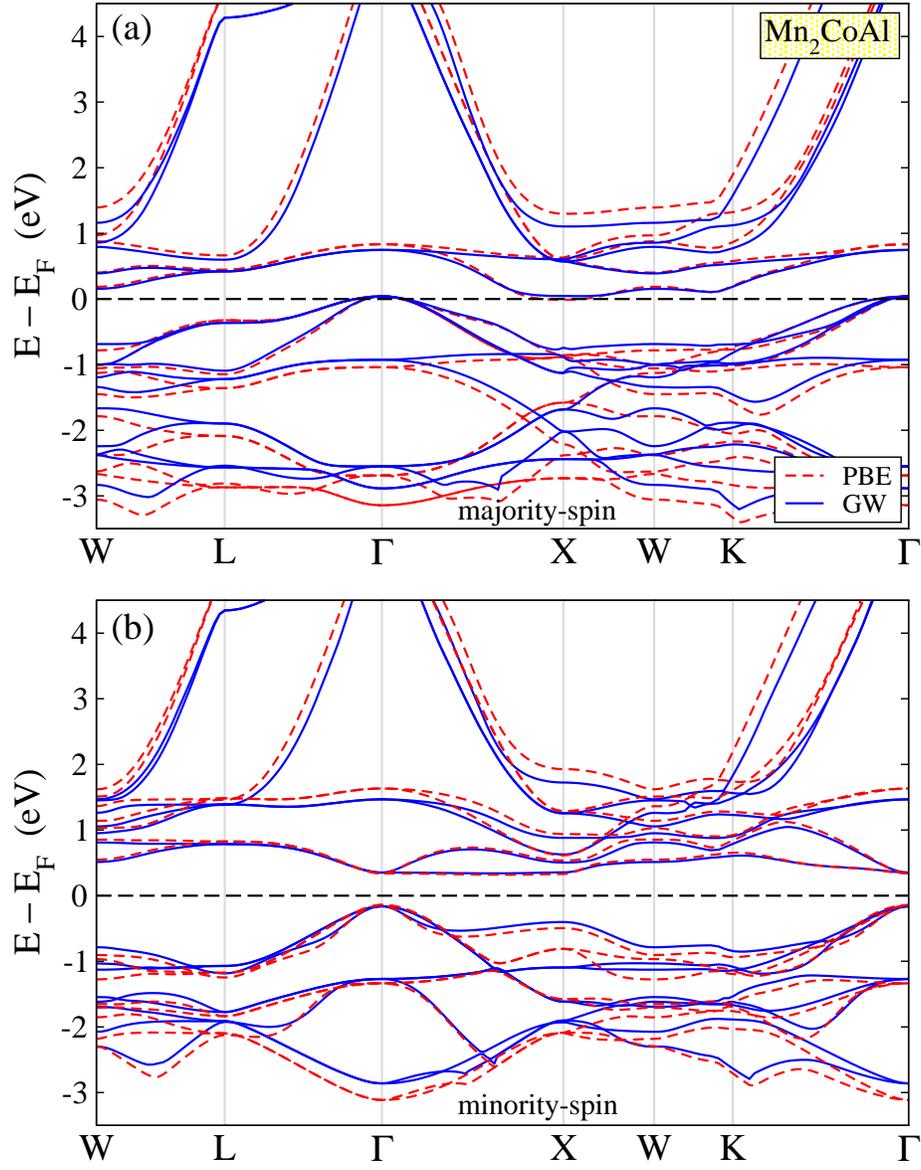}
\end{center} \caption{(Color online) Same as figure \ref{fig3} for Mn$_2$CoAl.
 \label{fig5}}
\end{figure}

\section{Results on spin-gapless semiconductors}

\begin{table*}
\caption{Same as table \ref{table2} for the spin-gapless
semiconductors.  \label{table3}}
\begin{tabular}{lccccccc}\hline
(XX$^\prime$)YZ &
E$_{\textrm{g}}^{GW}$(E$_{\textrm{g}}^\mathrm{PBE}$) & $\Gamma
\rightarrow \Gamma$ & X$\rightarrow$X & L$\rightarrow$L &
 $\Gamma \rightarrow$X & $\Gamma\rightarrow$L & X$\rightarrow$L \\
\hline
& \multicolumn{7}{c}{\underline{Majority-Spin Band-Structure}} \\
Cr$_2$ZnSi  &     0.00(0.00)&   3.23(3.27)&   1.21(1.32) &   0.00(0.00) &   2.01(1.81) &   1.71(1.67)&   0.91(1.17)\\
Mn$_2$CoAl  &     0.02(0.05)&   0.72(0.79)&   0.79(0.84) &   0.79(0.77) &   0.02(-0.05)&   0.39(0.40)&   1.17(1.30)\\
Ti$_2$CoSi  &     0.13(0.00)&   0.54(0.36)&   0.35(0.32) &   2.15(2.11) &   0.13(0.00) &   1.45(1.42)&   1.67(1.73)\\
Ti$_2$MnAl  &    -0.01(0.07)&   2.08(2.20)&   1.10(1.25) &   0.79(0.77) &   1.21(1.28) &   0.86(0.91)&   0.75(0.88)\\
& \multicolumn{7}{c}{\underline{Minority-Spin Band-Structure}} \\
Cr$_2$ZnSi &     0.31(0.44)&   2.60(2.94)&   0.71(1.08) &   0.31(0.44) &   1.50(1.67)&   1.45(1.65)&   0.67(1.06) \\
Mn$_2$CoAl &     0.53(0.47)&   0.52(0.48)&   0.77(0.82) &   1.86(2.00) &   0.53(0.47)&   0.96(0.97)&   1.20(1.32) \\
Ti$_2$CoSi &     0.92(0.79)&   1.69(1.42)&   2.06(2.05) &   1.80(1.72) &   1.35(1.20)&   1.16(0.97)&   1.86(1.83) \\
Ti$_2$MnAl &     0.67(0.61)&   1.29(1.09)&   1.70(1.30) &
1.76(1.64) &   0.67(0.63)&   0.73(0.60)&   1.75(1.28)
\\ \hline

\end{tabular}
\end{table*}

In the second part of our study, we focus on the SGS materials,
namely Cr$_2$ZnSi, Mn$_2$CoAl, Ti$_2$CoSi, and Ti$_2$MnAl. The
magnetic moments and the exchange splitting can be found in table
\ref{table1}.  The calculated magnetic energy (difference between
the total energy for a spin-polarized and a non-spin-polarized
calculation) suggests that these compounds prefer the magnetic
configuration and especially for Mn$_2$CoAl, which is the
prototype the value approaches -1 eV, and this explains why this
was the first SGS Heusler to be grown. Only Cr$_2$ZnSi has a very
small magnetic energy of -0.11 eV and thus would be more difficult
to stabilize its magnetic phase in experiments. The spin magnetic
moments have been already presented in detail in reference
\cite{Jakobsson}. Thus, we will only shortly discuss them here.
The most noticeable difference between this study and the one in
reference \cite{Jakobsson} is that for Mn$_2$CoAl we have used the
experimental lattice constant of 5.798 \AA , while in
\cite{Jakobsson} the equilibrium lattice constant of 5.73 \AA\ was
employed. The larger experimental lattice constant leads to spin
magnetic moments of larger absolute values, but the total spin
magnetic moment per formula unit remains equal to 2
$\mu_\mathrm{B}$ and Mn$_2$CoAl is a SGS for both lattice
constants. The exchange splitting values presented in the last
column of table \ref{table1} are of more interest. SGS materials
can also act as spin filters since they are just a special case of
magnetic semiconductors. With the exception of Ti$_2$CoSi, the
SGSs under study present exchange splitting energies  much larger
than the SFMs studied in the previous section. For Mn$_2$CoAl and
Ti$_2$MnAl the use of $GW$ instead of PBE has a minimal effect on
the obtained values contrary to the other two materials. For
Ti$_2$CoSi, $GW$ produces an almost vanishing exchange splitting,
while for Cr$_2$ZnSi $GW$ almost doubles the exchange splitting
with respect to the PBE.

In figure \ref{fig5} we present the spin-resolved band structure
for Mn$_2$CoAl  using both the PBE functional and the $GW$
approximation.  Contrary to (CoV)TiAl and (CrV)TiAl presented in
figures \ref{fig3} and \ref{fig4}, the situation for Mn$_2$CoAl is
completely different. Close to the Fermi level, $GW$ has a minimal
effect on the band structure for both spin directions, and hence
the energy band gaps and the transition energies presented in
table \ref{table3} are only marginally affected. The majority spin
band gap is 0.02 eV within $GW$ and 0.05 eV within PBE, and the
SGS character is conserved. In the minority spin band structure
the $GW$ calculated band gap is just 0.06 eV larger than the PBE
one. Similarly for the transition energies for Mn$_2$CoAl,  the
discrepancy between the PBE and $GW$ calculated values is
marginal, especially for the majority-spin band structure. The
other three compounds show a similar behavior. In the two
compounds containing Ti atoms (Ti$_2$CoSi and Ti$_2$MnAl) $GW$
slightly enlarges the band gaps and the transition energies,
similarly to the case of the SFMs. In the case of Cr$_2$ZnSi the
situation is opposite, and the $GW$ renormalization slightly
decreases the energy band gaps, especially for the minority-spin
band structure. Overall, in the case of the SGS materials, the
$GW$ self-energy has only a small effect on the electronic band
structure, but it may affect the exchange splittings significantly
with consequences for spin-filter related applications.

\section{Conclusions}

The $GW$ approximation for the electronic self-energy was employed
to account for many-body exchange-correlation effects as a
correction to standard density-functional theory calculations
using the PBE functional. We have studied the properties of two
distinct subfamilies of the full-Heusler compounds: (i) the
ordered quaternary (also known as LiMgPdSn-type) Heusler compounds
having the chemical formula (CoV)YAl and (CrV)YAl,  where Y is Ti
or Zr, and which are magnetic semiconductors (known as spin-filter
materials), and (ii) the so-called inverse Heusler compounds with
the chemical formula X$_2$YZ which are spin-gapless semiconductors
and thus present a gapless or almost-gapless semiconducting
behavior in the majority spin band structure combined with a
finite energy gap in the other spin channel.

Our first-principles results suggest that the use of $GW$ is
important for the spin-filter materials. It  shifts both valence
and conduction bands away from the Fermi level leading to larger
energy band gap values as well as larger transition energies. On
the other hand, the effect of employing $GW$  is smaller in the
spin-gapless semiconductors and the usual density-functional
theory gives a fair description of the properties of these
materials.

Thus the effect of the $GW$ approximation is material-specific
even among materials of the same family with similar electronic
and magnetic properties and its use seems essential to get a good
description of their electronic properties. We hope that our
results further enhance the interest in these classes of Heusler
compounds and that they contribute to the understanding of their
extraordinary properties.

\end{document}